\documentclass[journal]{IEEEtran}
\usepackage{cite}
\usepackage[pdftex]{graphicx}
\usepackage[pdftex]{color}
\graphicspath{/}
\usepackage[ruled,vlined,linesnumbered]{algorithm2e}
\usepackage[cmex10]{amsmath}
\usepackage{algpseudocode}
\usepackage{array}
\usepackage{url}
\usepackage{mathtools}
\usepackage{color}
\usepackage{amsmath, amssymb}

\usepackage{amsthm}

\theoremstyle{remark}

\theoremstyle{definition}

\begin{document}
\title{Reliable Reporting for Massive M2M Communications with Periodic Resource Pooling}

\author{Germ\'an Corrales Madue\~no, Cedomir Stefanovi\' c, Petar Popovski
\thanks{The authors are with the Department of Electronic Systems, Aalborg University, Aalborg, Denmark (e-mail: \{gco, cs, petarp\}@es.aau.dk).}
\thanks{The research presented in this paper was supported by the Danish Council for Independent Research (Det Frie Forskningsr\aa d) within the Sapere Aude Research Leader program, Grant No. 11-105159 ``Dependable Wireless~bits for Machine-to-Machine (M2M) Communications''.}
}

\maketitle

\begin{abstract}
This letter considers a wireless M2M communication scenario with a massive number of M2M devices. 
Each device needs to send its reports within a given deadline and with certain reliability, e. g. $99.99$\%. A pool of resources available to all M2M devices is periodically available for transmission. The number of transmissions required by an M2M device within the pool is random due to two reasons - random number of arrived reports since the last reporting opportunity and requests for retransmission due to random channel errors. We show how to dimension the pool of M2M-dedicated resources in order to guarantee the desired reliability of the report delivery within the deadline. The fact that the pool of resources is used by a massive number of devices allows to base the dimensioning on the central limit theorem. The results are interpreted in the context of LTE, but they are applicable to any M2M communication system.  
\end{abstract}


\IEEEpeerreviewmaketitle

\section{Introduction}
\label{sec:intro}
	
In the past two decades, the main focus of the cellular access engineering was on the efficient support of human-oriented services, like voice calls, messaging, web browsing and video streaming services.
A common feature of these services is seen in the relatively low number of simultaneous connections that require high data rates.
On the other hand, the recent rise of M2M communications introduced a paradigm shift and brought into research focus fundamentally new challenges.
Particularly, M2M communications involve a massive number of low-rate connections, which is a rather new operating mode, not originally considered in the cellular radio access.

Smart metering is a showcase M2M application consisting of a massive number of devices, up to $30 000$\cite{smartgrid}, where meters periodically report energy consumption to a remote server for control and billing purposes. On the other hand, the report size is small, of the order of $100$ bytes \cite{3GPP_Smart}.
The current cellular access mechanisms, considering all the associated overhead, can not support this kind of operation \cite{somecitations1}. 
There are on-going efforts in 3GPP that deal with the cellular access limitations, investigating methods for decreasing the access overhead for small data transmissions \cite{23.887}, access overload control \cite{23.888} and guaranteed quality of service \cite{QoSLTE}. Besides LTE, \cite{ICC13} proposes an allocation method for reports with deadlines in GPRS/EDGE, showing that up to $10^4$ devices can be effectively supported in a single cell by avoiding random access and using a periodic structure to serve the devices such that the deadlines are met. 

In this letter, we consider a system with a periodically occurring pool of resources that are reserved M2M communications and shared for uplink transmission by all M2M devices. The re-occurring period is selected such that if a report is transmitted successfully within the upcoming resource pool, then the reporting deadline is met.
We note that, if each device has a deterministic number of packets to transmit in each resource pool and if there are no packet errors, then the problem is trivial, because a fixed number of resources can be pre-allocated periodically to each device. However, if the number of packets, accumulated between two reporting instances, is random and the probability of packet error is not zero, then the number of transmission resources required per device in each transmission period is random. On the other hand, as the number of transmission resources in each instance of the resource pool is fixed, the following question arises: \emph{How many periodically reporting devices can be supported with a desired reliability of report delivery (i.e., $99.99$\%), for a given number of resources reserved for M2M communications?} We elaborate and analyze the proposed approach in LTE context; however, the presented ideas are generic and implementable in other systems where many devices report to a single base station or access point. 

The rest of the letter is organized as follows.
	Section~\ref{sec:model} presents the system model.
	Section~\ref{sec:analysis} is devoted to the analysis of the proposed access method.
	Section~\ref{sec:results} presents the numerical results and Section~\ref{sec:conclusions} concludes the letter.

\section{System Model}
\label{sec:model}
	
	We focus on the case of periodic reporting, where the length of the reporting interval (RI), denoted by $T_{RI}$, depends on the application requirements \cite{3GPP_Smart}. The M2M resources for uplink transmission are reserved to occur periodically, at the end of each RI. The periodic reporting is typically modeled either as a Poisson process with arrival rate $\lambda = 1 / T_{RI}$, where devices can actually send none, one, or multiple reports within RI, or as a uniform distribution, where devices send a single packet per RI \cite{IEEE802.16p0014,3GPP_USF}. Our analysis will focus on the former case, but we note that the derived results can be readily applied to the latter arrival model, as well. We assume that all report arrivals that occur within the current reporting interval are served in the next reporting interval. 
	
	\begin{figure}
	  \centering
	    \includegraphics[width=0.95\columnwidth]{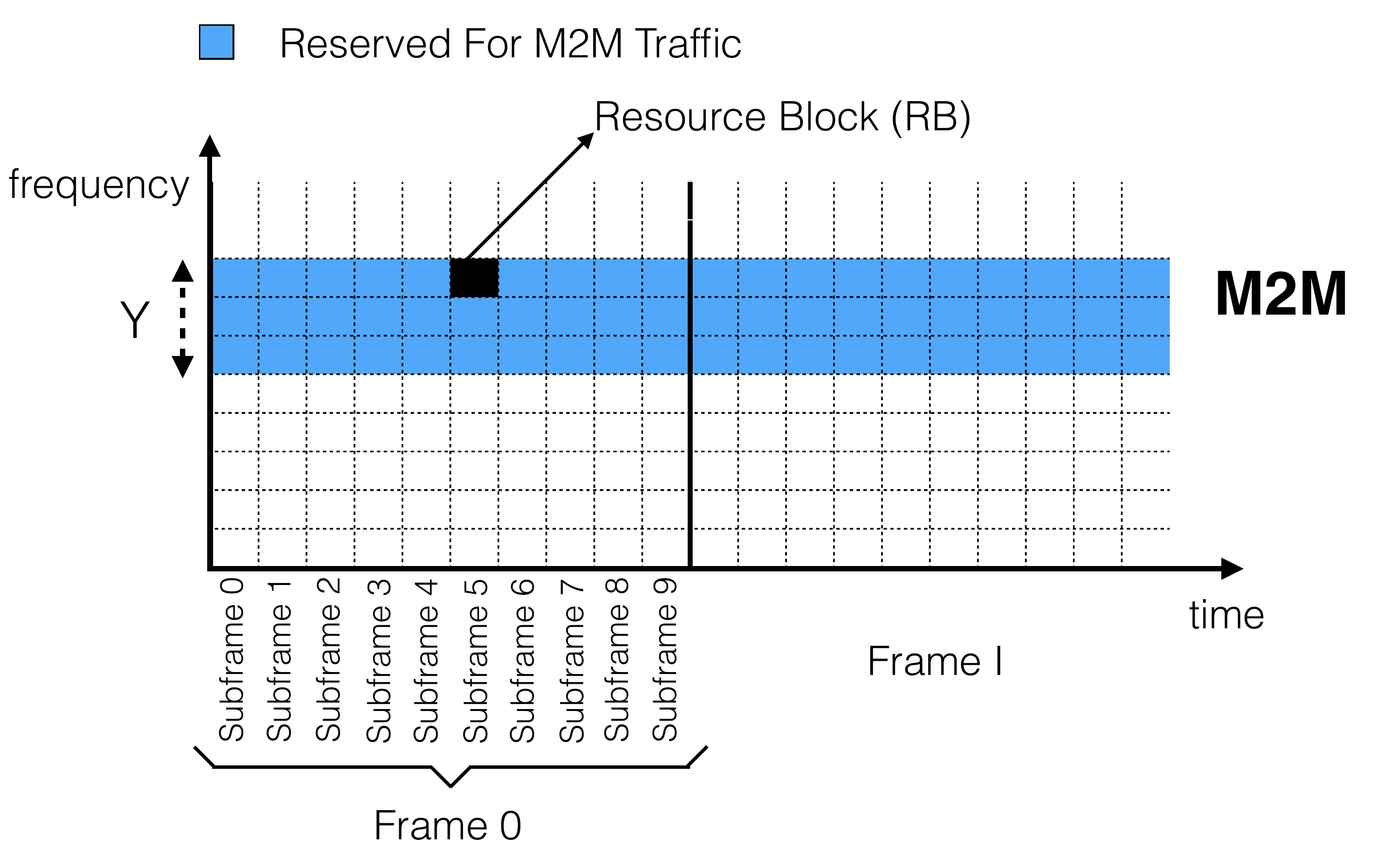}\caption{Representation of the LTE uplink resource structure, where a set of RBs has been reserved for M2M purposes.}\label{LTE_grid}
	    \vspace{-12pt}
	\end{figure}
	
The LTE access combines TDMA and FDMA, such that access resources are represented in 2D, see Fig.~\ref{LTE_grid}.
As depicted, time is organized in frames, where each frame is composed of subframes with duration $T_s = 1$~ms.
The minimum amount of uplink resources that can be allocated to a device is one resource block (RB), corresponding to a single subframe and 12-subcarriers in frequency.
We assume that the uplink resources are split into two pools, one reserved for M2M services (depicted in blue in Fig.~\ref{LTE_grid}), and the other used for other services. Note that the approach of splitting the resources for M2M and non-M2M has often been used \cite{preambles,paviaAGCH}, as their requirements are fundamentally different.
Finally, we assume that there is a set of $Y$ RBs reserved for M2M resource pool in each subframe. 
		
\begin{figure}
	  \centering
	    \includegraphics[width=0.75\columnwidth]{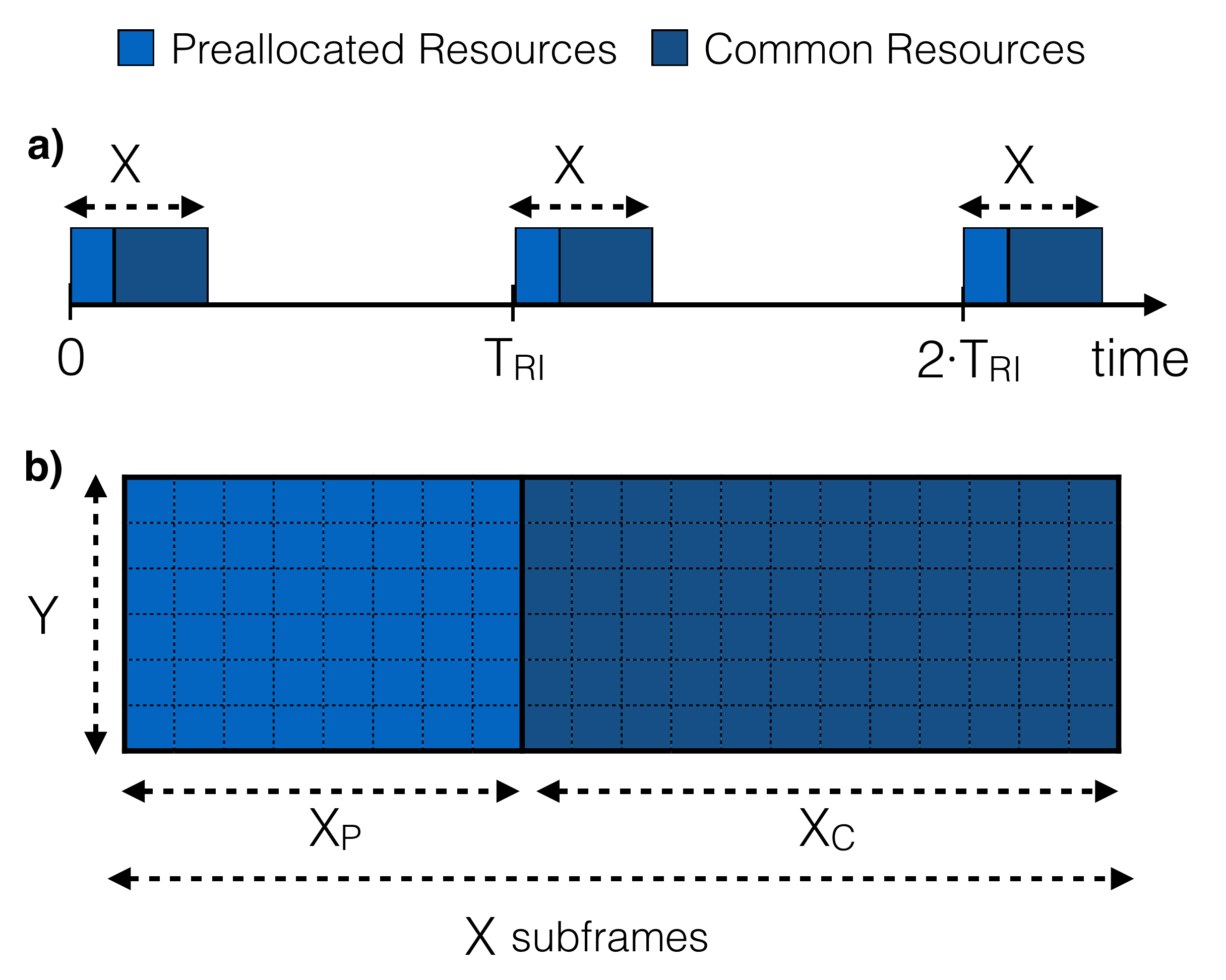}\caption{a) Periodically occurring M2M resource pool. b) Division of M2M resource pool in the pre-allocated and common pool.}\label{preallocated}
	    \vspace{-12pt}
	\end{figure}

	The M2M resource pool is divided into two parts, denoted as the preallocated and common pool, which reoccur with period $T_{RI}$, as depicted in Fig.~\ref{preallocated}a). We assume that there are $N$ reporting devices, and each device is preallocated an amount of RBs from the preallocated pool dimensioned to accommodate a single report and an indication if there are more reports, termed excess reports, from the same device to be transmitted within the same interval.
The common pool is used to allocate resources for the excess reports, as well as all the retransmissions of the reports/packets that were erroneously received. These resources can only be reactively allocated and in our case we consider the LTE data transmission scheme, where each transmission has an associated feedback that can be used to allocate the resources from the common pool\footnote{The minimum latency for the feedback is $6$~ms ($6$ subframes), which includes processing times at the base station and at the device, and which can be assumed negligible taking into account that the RI that we are considering is of the order of thousands subframes.}. The length of the M2M resource pool, preallocated  pool and common pool, expressed in number of subframes, are denoted by $X$, $X_P$ and $X_C$, respectively, see Fig.~\ref{preallocated}b), such that 
\begin{align}
	X = X_P + X_C = \alpha N + X_C.
\end{align}	
where $\alpha \leq 1$ denotes the fraction of RBs per subframe required to accommodate a report transmission and where the value of $X_C$ should be chosen such that a report is served with a required reliability. The analysis how to determine $X_C$, given the constraints of the required number of RBs per report, number of devices and reliability, is the pivotal contribution of the letter and presented in Section~\ref{sec:analysis}. Finally, we note that the duration of $X$ has a direct impact on the delay; in the worst case a (successful) report is delivered $T_{RI}+(X\cdot T_s)$ seconds after its arrival, which also defines the delivery deadline.

\section{Analysis}
\label{sec:analysis}		

	
	Denote by $W_{i,j}$ a random variable that models the total number of transmissions of report $j$ from device $i$; i.e., $W_{i,j}$ includes the first transmission and any subsequent retransmissions that may occur due to reception failures. Further, assume that the maximum value of $W_{i,j}$ is limited to $L$, where $L$ is a system parameter, applicable to all reports from all devices.
	The probability mass function (pmf) of $W_{i,j}$ can be modeled as a geometric distribution truncated at $L$: 
	\begin{align}
	  P[W_{i,j}=k]=\left\{
		  \begin{array}{l l}
		    p_e^{k-1}  (1-p_e), &  1 \leq k \leq L-1, \\
		  	 p_e^{L-1}, & k = L,
		  \end{array} \right. \label{h0}	
	\end{align}
	where $p_e$ is the probability of a reception failure.
	
	Recall that the reporting of device $i$ is modeled as a Poisson process $U_i$ with arrival rate $\lambda$, where the device can send none ($U_i=0$), one ($U_i=1$) or multiple reports ($U_i \geq 1$) within RI.
	As stated above, the first transmission of the first report of a device is sent in the preallocated pool, while all subsequent transmissions take place in the common pool.
	These include: a) potential retransmissions of the first report, and b) transmissions and potential retransmissions of all excess reports.
	Denote by $R_i$ the random variable that corresponds to the total number of transmissions from device $i$ requiring resources in the common pool:
	\begin{align}
	R_i = \left\{
	\begin{array}{l l}
		    0, &  U_i = 0, \\
		  	\sum_{j=1}^{U_i} W_{i,j} - 1, & U_i \geq 1.
		  \end{array} \right.
	\label{eq:R_i}
	\end{align}
	The total number of transmissions of all devices requiring resources from the common pool is:
	\begin{align}
	R = \sum_{i=1}^{N} R_i.
	\end{align}
	The straightforward way to characterize $R$ is to derive its pmf.
	However, as the supposed number of reporting devices is very large, $R$ is a sum of a large number of independent identically distributed (IID) random variables.
	Therefore, we apply the central limit theorem and assume that $R$ is a Gaussian random variable, requiring 
characterization only in terms of the expectation and variance.
We proceed by evaluation of $\mathrm{E}[R]$ and $\mathrm{\sigma}^2[R]$, and show in Section~\ref{sec:results} that this approach provides accurate results.

	\subsection{Expectation and variance of $R$}
	
		The expectation of $R$ is:
		\begin{equation}
			\mathrm{E} [ R ] = \mathrm{E}\left[\sum_{i=1}^N R_i\right] = N \cdot \mathrm{E}\left[R_i\right].
			\label{ER}
		\end{equation}
		Taking into account \eqref{eq:R_i}, it could be shown that:
		\begin{align}
		\mathrm{E} [ R_i ] & = \mathrm{E} [ R_i | U_i = 0 ] \mathrm{P} [ U_i = 0 ] + \mathrm{E} [ R_i | U_i \geq 1 ] \mathrm{P} [ U_i \geq 1 ], \nonumber \\
											 & = \mathrm{E} [ R_i | U_i \geq 1 ] \mathrm{P} [ U_i \geq 1 ], \nonumber \\
											 & = \mathrm{E} \left[ \sum_{j=0}^{U_i} W_{i,j} - 1 | U_i \geq 1 \right] ( 1 - e^{-\lambda T_{RI}}) \label{eq:exp}.								 
		\end{align}
		By putting $\lambda T_{RI} = 1$ and applying Wald's equation \cite{randomSum}, the equation \eqref{eq:exp} becomes:
		\begin{align}									 
		\mathrm{E} [ R_i ] & = \left( \mathrm{E} [ U_i | U_i \geq 1 ] \mathrm{E} [ W_{i,j} ] - 1 \right) ( 1 - e^{ - 1 }), \nonumber \\
											 & = \frac{1 - p_e^L }{ 1 - p_e } - ( 1 - e^{ - 1 }). \label{eq:exp1}
		\end{align}
		where we used the fact that $\mathrm{E} [ U_i | U_i \geq 1 ] = 1 / ( 1 - e^{ - 1 } )$.
		Substituting \eqref{eq:exp1} into \eqref{ER} yields expectation of $R$.
	
		The variance of $R$ can be derived in a similar fashion, using the identities related to the variance of the random sum of random variables \cite{randomSum}.
		Due to the space limitations, we omit the derivation and present only the final result:
		\begin{align}
			\mathrm{\sigma^2}\left[R\right] &= N \left[ \frac{(2L-1) p_e^{L+1} - (2L+1) p_e^{L} + p_e + 1}{(1-p_e)^2} + \right. \nonumber \\
			& \left.  +  e^{-1} \cdot \left( 1 - 2 \cdot \frac{1-p_e^L}{1-p_e} - e^{-1} \right) \right] .
		\end{align}

	\subsection{Probability of Report Failure }
		
		In this subsection, we assess the probability of report failure, i.e., the probability that the report has not been successfully delivered after all attempted (re)transmissions.
		In general, this probability depends both on the number of resources available and the scheduling policy applied in the common pool.
		In order to avoid the particularities related to scheduling, we derive an upper bound on the probability of failure which is valid for any scheduling policy.
		
		Denote by $\Phi$ the event that a report experiences a failure.
		Further, denote by $l$ the number of required report transmissions, which includes the first transmission and all the required retransmissions.
		If we assume for a moment that the number of available resources in the common pool is infinite, then the report fails to be delivered only when the required number of transmissions exceeds $L$: 
		\begin{equation}	
			\mathrm{P}_{\infty}[ \Phi ] = \mathrm{P}_{\infty}[ \Phi, l > L ] = \mathrm{P} [ l > L ] = p_e^L. 
		\end{equation}
If the common pool that consists of $X_C$ subframes can accommodate $C$ transmissions (i.e., $C$ is the capacity of the common pool in number of transmissions), then the probability of report failure can be written as:
		\begin{align}
			\mathrm{P}[ \Phi ] &  = \sum_{k = 1}^{L} \mathrm{P} [ \Phi , l = k ] + \mathrm{P} [ \Phi , l > L ], \nonumber \\
												 & = \sum_{k = 1}^{L} \mathrm{P} [ \Phi | l = k ] P [ l = k ] + p_e^L . \label{eq:phi}
		\end{align}
		Further, for $1 \leq k \leq L$:
		\begin{align}
			\mathrm{P} [ \Phi | l = k ] & = \mathrm{P} [ \Phi, R > C | l = k ] + \mathrm{P} [ \Phi, R \leq C | l = k ], \nonumber \\
			 														& = \mathrm{P} [ R > C ] \mathrm{P} [ \Phi | l = k, R > C ],	 	\label{eq:phi_sched}											
		\end{align}
		where we used the fact that $\mathrm{P} [ \Phi, R \leq C | l = k ] = 0$, i.e., there is no report delivery failure when the total number of required transmissions $R$ is not greater then the capacity of the common pool $ C $, for $l \leq k \leq L$.
Regardless of the scheduling policy applied in the common pool, it is always $\mathrm{P} [ \Phi | l = k, R > C ] \leq 1$, which leads to the following upper bound:
\begin{align}
\mathrm{P} [ \Phi | l = k ] \leq \mathrm{P} [ R > C ]. \label{eq:bound}
\end{align}				
Finally, substituting \eqref{eq:bound} into \eqref{eq:phi} yields:
		\begin{align}
			P[ \Phi ] \leq  Q\left( \frac{C - \mu}{\sigma} \right)  (1-p_e^L) + p_e^L, \label{eq:q}
		\end{align}
where $\mu = \mathrm{E}[R]$, $\sigma = \sqrt{\sigma^{2}[R]}$, and $Q( \cdot )$ is Q-function, standardly used for Gaussian random variables.

\section{Results}
\label{sec:results}	
	
\begin{figure}
	  \centering
	    \includegraphics[width=\columnwidth]{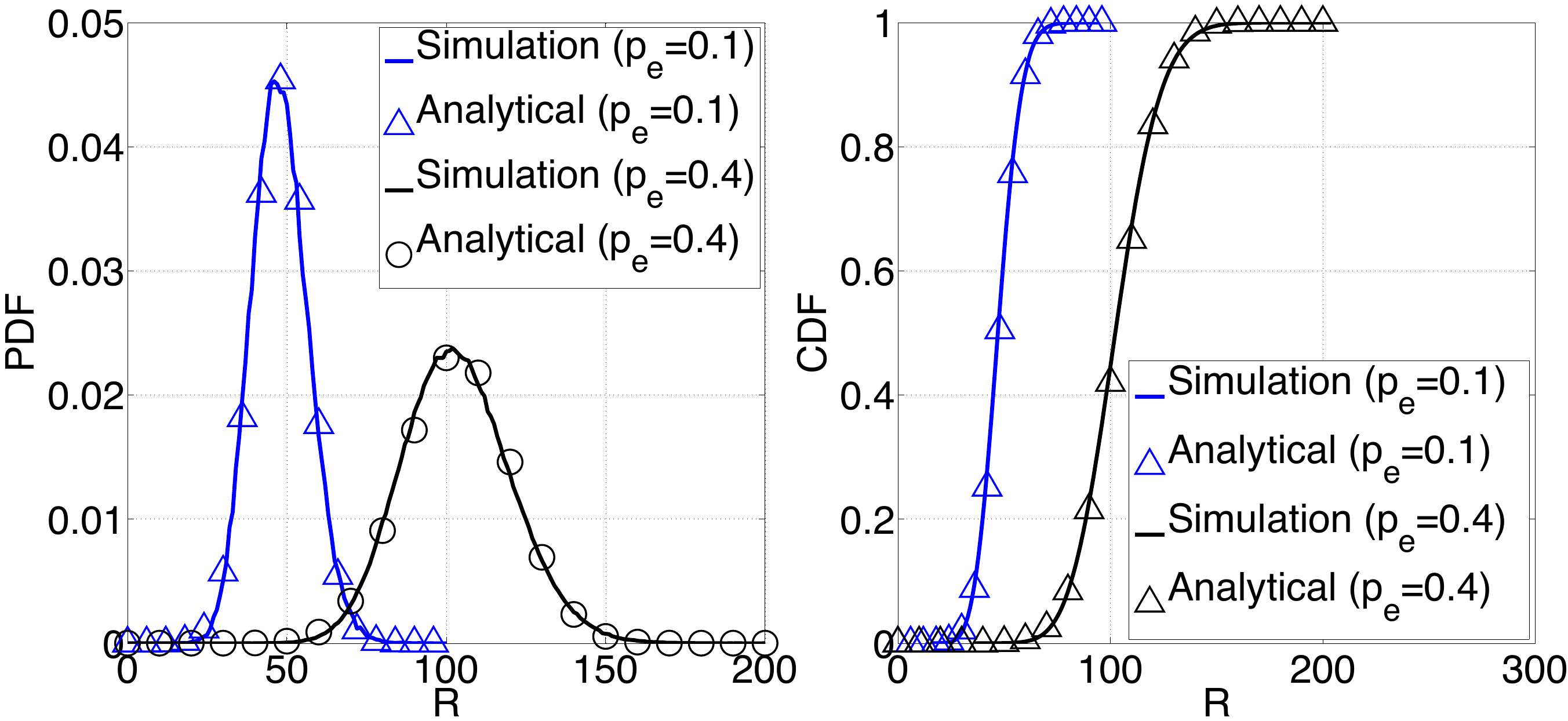}\caption{Comparison of simulated and analytical pdf and cdf of $R$ when $N=100$ and $L=10$.}\label{validation}
	    \vspace{-12pt}
	\end{figure}

We first validate the Gaussian assumption in the analysis by comparing the probability density function (pdf) and cumulative density function (cdf) of the Gaussian model with the simulated ones. Fig.~\ref{validation} presents a tight match between the model and simulations, when number of reporting devices is $N=100$, the maximum number of transmissions per report is $L=10$, the probability of report error $p_e$ takes values $0.1$ and $0.4$, respectively, and the number of the simulation runs is set to $10^5$ for each parameter combination.

	Further, using the bound derived in \eqref{eq:q}, we determine the percentage of LTE system resources required for reliable M2M services, defined as the ratio of RBs required for M2M and the total amount of RBs available in the system (see Fig.~\ref{LTE_grid}).
	LTE foresees a maximum bandwidth of 100~RBs (i.e., 20~MHz) per subframe, see Fig.~\ref{LTE_grid}, while 20~RBs per subframe (i.e., 5~MHz) are most commonly used \cite{5MHz}.
	The amount of RBs required to report transmission depends on the used modulation\footnote{In this work we consider the lowest-order (QPSK) and the highest-order (64-QAM) LTE modulation schemes.} and the report size (RS).
	For the individual transmission error we used $p_e=0.1$, which is a typical value for LTE data transmissions \cite{ahmadi2013lte}, while the maximum number of report transmissions is again set to $L=10$.
	The maximal number of devices is set to $30000$, as indicated by 3GPP in \cite{smartgrid}. 
	Finally, we set the probability of report failure to $P[\Phi] \leq 10^{-3}$, i.e. the desired reliability to at least $99.99$\%,
	To validate the analytical upper bound, we simulated a random scheduler with $10^5$ repeats for each parameter combination.
	
\begin{figure}
	  \centering
	    \includegraphics[width=0.85\columnwidth]{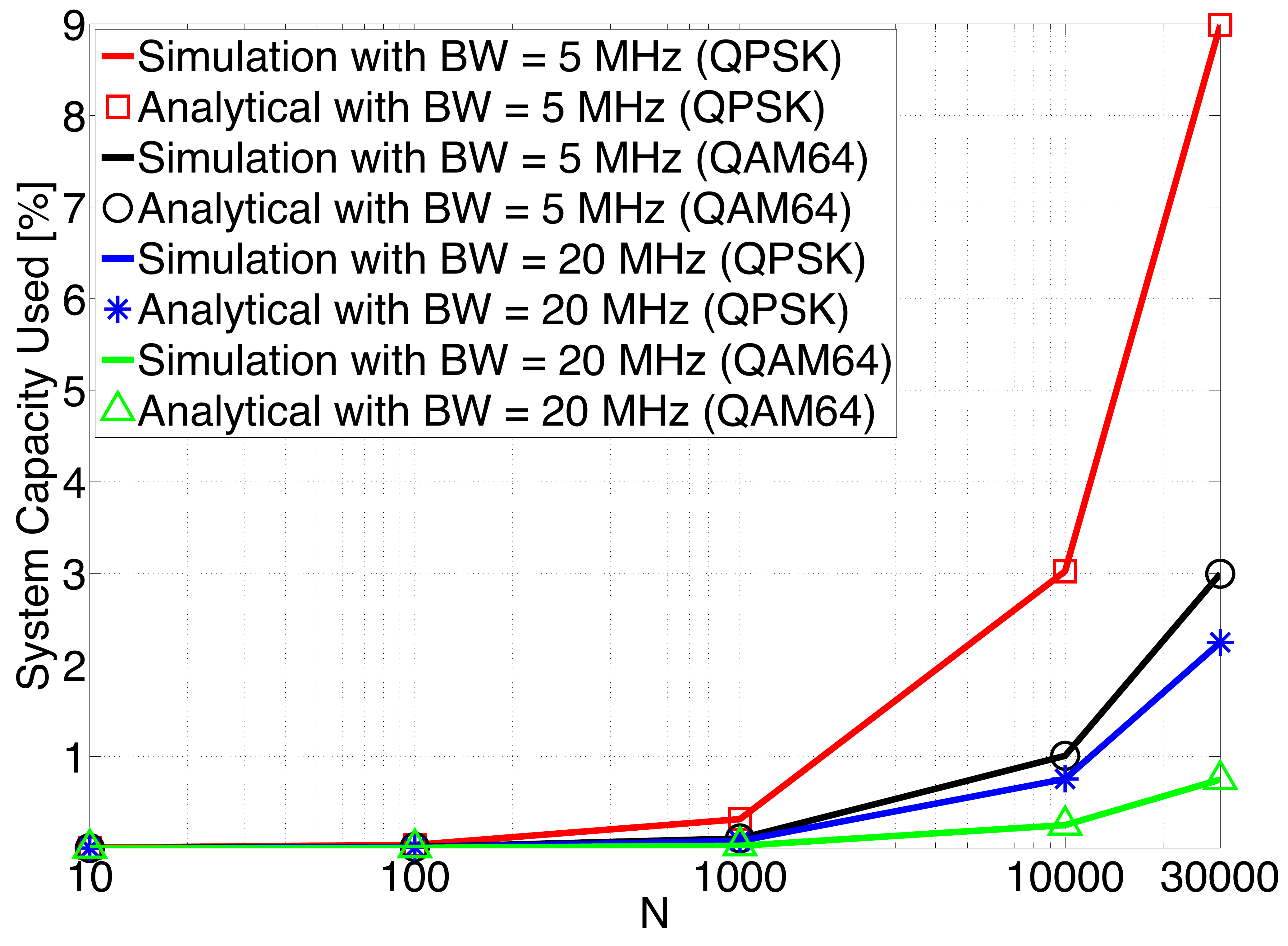}\caption{Fraction of system capacity used for M2M services, when $P[\Phi] \leq 10^{-3}$, RI of 1~minute, RS of 100~bytes and $p_e=10^{-1}$.}\label{Fig1}
	\end{figure}

	Fig.~\ref{Fig1} shows the percentage of system resources required to serve $N$ devices for a reporting interval RI of $1$~minute, which corresponds to the most demanding RI according to \cite{3GPP_Smart}. The report size (RS) is $100$~bytes.
	It can be seen that in the worst case, for the available bandwidth of $5$~MHz and the lowest-order modulation (QPSK), up to $30000$ devices can be served with only $9$\% of the available system resources. If a larger bandwidth ($5$~MHz) and/or higher-order modulations ($64$-QAM) is used, then only $3$\% of the available resources are required to achieve the target report reliability.
	
\begin{figure}
	  \centering
	    \includegraphics[width=0.85\columnwidth]{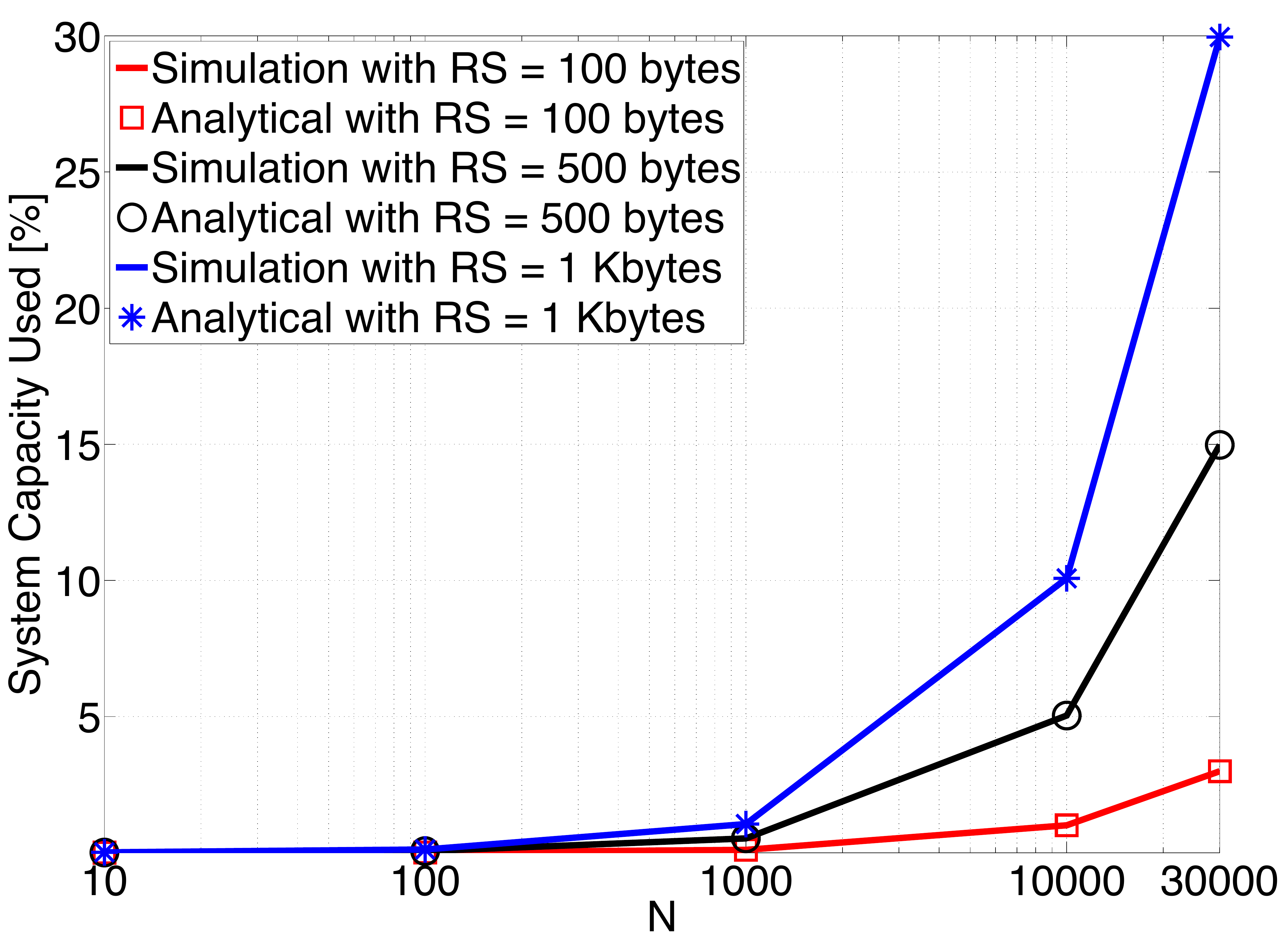}\caption{Fraction of system capacity used for M2M services, when $P[\Phi] \leq 10^{-3}$, RI of 1~minute, bandwidth of 5~MHz, 64-QAM and $p_e=10^{-1}$.}\label{Fig3}
	    \vspace{-12pt}
	\end{figure}

	Fig.~\ref{Fig3} depicts the required fraction of system capacity for M2M service, when the report RS varies between $100$~bytes and $1$~kbytes\cite{3GPP_Smart}, the system bandwidth is set to $5$~MHz, modulation scheme is $64$-QAM, and $p_e=0.1$. Obviously, the report size has a large impact in the results, demanding up to 30\% of the system capacity in the worst case.
	
	Finally, we note that Figs.~\ref{Fig1} and \ref{Fig3} also demonstrate a tight match between the analytical and simulated results.
	
\section{Conclusions}
\label{sec:conclusions}

We have introduced a contention-free allocation method for M2M that relies on a pool of resources reoccurring periodically in time. Within each occurrence, feedback is used to reactively allocate resources to each individual device. The number of transmissions required by an M2M device within the pool is random due to two reasons: (1) random number of arrived reports since the last reporting opportunity and (2) requests for retransmission due to random channel errors. The objective is to dimension the pool of M2M-dedicated resources in order to guarantee certain reliability in the delivery of a report within the deadline. The fact that the pool of resources is used by a massive number of devices allows to base the dimensioning on the central limit theorem.  
Promising results have been shown in the context of LTE, where even with the lowest-order modulation only $9$\% of the system resources are required to serve $30$K M2M devices with a reliability of $99.99$\% for a report size of $100$~bytes. The proposed method can be applied to other systems, such as $802.11$ah.		
	
\end{document}